\documentclass{IEEEtran}
\usepackage{cite}
\usepackage{amsmath,amssymb,amsfonts}
\usepackage{graphicx}
\usepackage{textcomp,nicefrac}
\usepackage{hyperref}
\def\BibTeX{{\rm B\kern-.05em{\sc i\kern-.025em b}\kern-.08em
T\kern-.1667em\lower.7ex\hbox{E}\kern-.125emX}}
\begin{document}
\title{Measurement of Position Resolutions of L-band Cavity Beam Position Monitors}
\author{Soohyung Lee, 
Ho Jun Jeong,
JongMo Hwang,
GwangUk Park,
Siwon Jang,
Konstantin Popov,
Alexander Aryshev,
Toshiyuki Okugi,
Eun San Kim
\thanks{This work was supported by the National Research Foundation of Korea (NRF) grant
funded by the Korea government (MSIT) (RS-2022-00143178, RS-2025-23963275) and
"MEXT Development of key element technologies to improve the performance of future
accelerators Program", Japan Grant Number JPMXP1423812204.}
\thanks{S. Lee is with the Center for Accelerator Research, Korea University, Sejong, 30019, Republic of Korea (e-mail: huvx44@korea.ac.kr).}
\thanks{H. J. Jeong and J. Hwang are with the Department of Accelerator Science, Korea University, Sejong, 30019, Republic of Korea.}
\thanks{G. Park is with the Division of Semiconductor Physics, Korea University, Sejong, 30019, Republic of Korea.}
\thanks{S. Jang is with the PLS-II Accelerator Department, Pohang Accelerator Laboratory, Pohang, 37673, Republic of Korea.}
\thanks{K. Popov, A. Aryshev and T. Okugi are with the Innovation Center for Applied Superconducting Accelerators, High Energy Accelerator Research Organization (KEK), Tsukuba, Ibaraki 305-0801, Japan.}
\thanks{E. S. Kim is with the Department of Accelerator Science and the Center for Accelerator Research, Korea University, Sejong, 30019, Republic of Korea (e-mail: eskim1@korea.ac.kr).}
}

\maketitle

\begin{abstract}
    Beam position monitors (BPMs) are indispensable components of modern particle accelerators, 
    providing real-time diagnostics to ensure precise beam control, stability, and quality. 
    As accelerators such as the International Linear Collider (ILC) aim for nanometer-scale beam sizes 
    at the interaction point, stringent requirements on position resolution arise. Specifically, 
    the main linac of the ILC demands a BPM resolution better than 5\,$\mu$m to support stable beam transport 
    and minimize emittance growth. To address this, we have developed an L-band cavity BPM 
    optimized for the beam conditions of the ILC. 
    In this paper, we introduce a prototype of an L-band cavity BPM and its signal processing system,
    describe the methodology for position resolution measurements, discuss the problems and solutions
    encountered in the past experiment, and report the
    projected position resolutions of about 300\,nm at best.
\end{abstract}


\section{Introduction}
\label{sec:introduction}
Modern particle accelerators are crucial tools for exploring fundamental physics to unveil the mysteries of the universe.
High-energy particle accelerators have been enabling triumphs of discoveries such as 
the Higgs boson at the Large Hadron Collider (LHC) \cite{higgs_discovery}.
There are yet remaining questions in particle physics that demand the next generation accelerators with higher energy and luminosity
such as the International Linear Collider (ILC) \cite{ilc_tdr}, the Future Circular Collider (FCC) \cite{fcc_cdr}, 
and the Circular Electron Positron Collider (CEPC) \cite{cepc_cdr}. These future accelerators require extremely small beam sizes 
at the interaction point to achieve high luminosity. For example, the ILC aims to achieve beam sizes of 5.9\,nm 
in the vertical and 474\,nm in the horizontal at the interaction point \cite{ilc_tdr}. 
To maintain such small beam sizes throughout the accelerators, precise beam position monitorings and controls are essential.

Beam position monitors (BPMs) are critical components in particle accelerators, providing real-time diagnostics 
of the beam position and trajectory. There are various types of BPMs in the literature, and cavity BPMs are known to provide
the highest position resolution among them, for example, a few nanometers of position resolution is achievable for the ILC
interaction point \cite{prst_11_062801}.

In this paper, we introduce a prototype of an L-band cavity BPM dedicated for the main linac of the ILC. 
Since the most important factor of a BPM is its position resolution, we focus on the methodology 
and results of the position resolution measurements of the developed BPMs.

\begin{figure}[b]
\centerline{\includegraphics[width=3.5in]{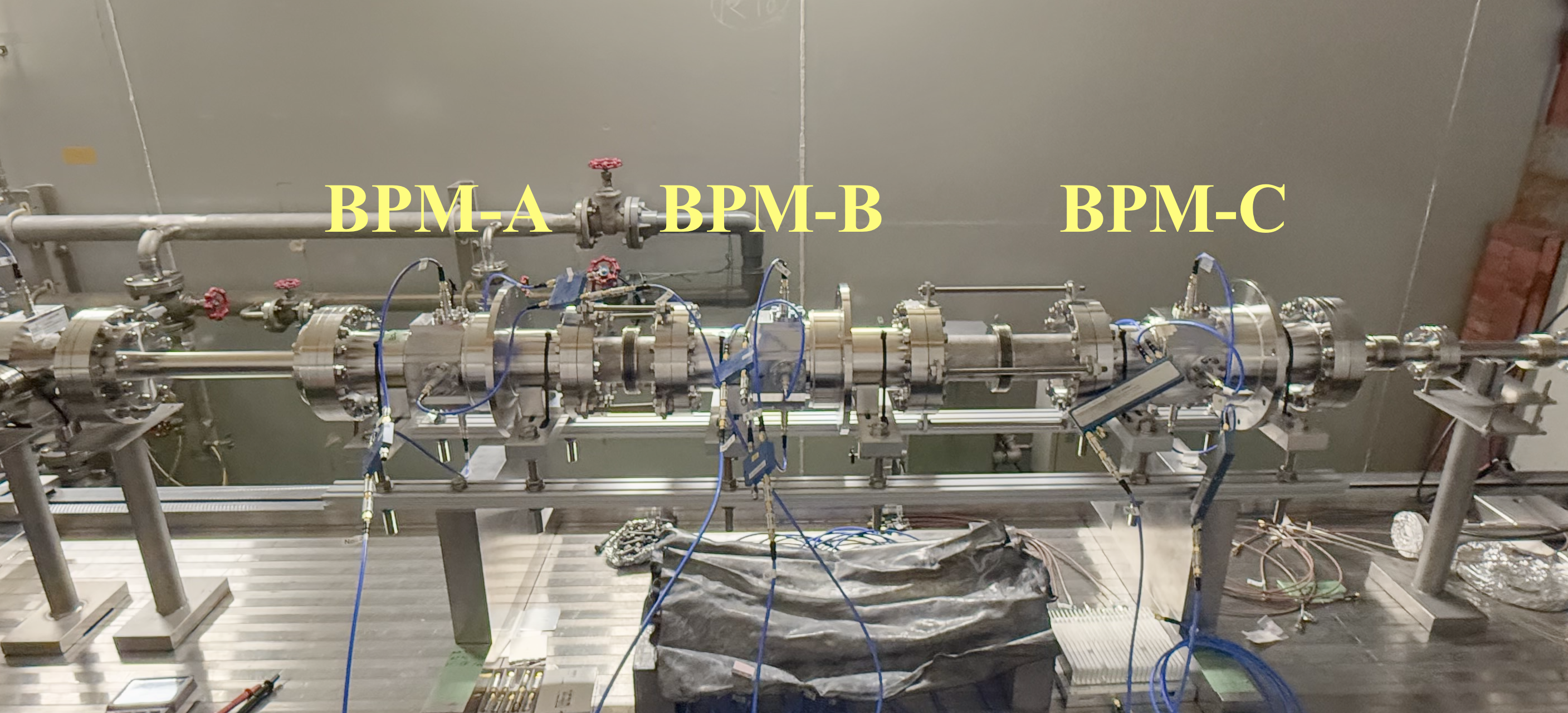}}
\caption{\label{fig:bpm_station}
Three L-band cavity BPMs installed in the ATF. The beam enters from the left side, and the cavity BPMs are labeled
as BPM-A, BPM-B, and BPM-C from upstream to downstream. The electronics for the signal processing are placed
under the BPM-B covered by a black lead sheet and lead blocks for a radiation shielding.}
\end{figure}

\begin{figure*}[t]
\centerline{\includegraphics[width=7in]{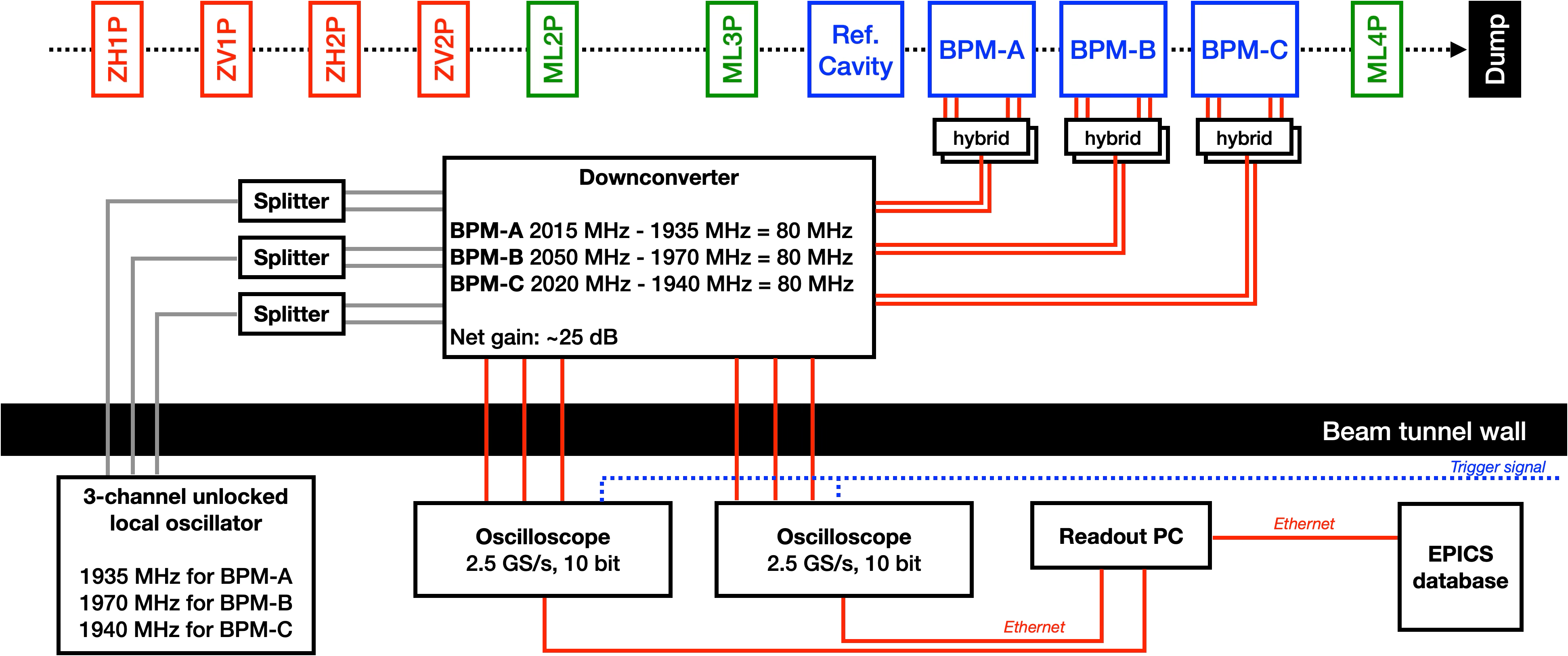}} 
\caption{\label{fig:rf_chain}
The signal processing chain for the L-band cavity BPMs illustrated along with other components of the ATF such as
the steering magnets and stripline BPMs. The red rectangles (labeled as ``ZH1P'', ``ZV1P'', ``ZH2P'', and ``ZV2P'') indicate the
steering magnets (H and V for horizontal and vertical, respectively) and the green rectangles (labeled as ``ML2P'', ``ML3P'',
and ``ML4P'') indicate the stripline BPMs. The blue rectangles indicate cavity BPMs. The first cavity BPM labeled as
``Ref. Cavity'' is a reference cavity which was not used in this work. The downconverter is located inside the beam tunnel whereas 
the LO source and the oscilloscopes are placed outside the tunnel.}
\end{figure*}

\section{L-band Cavity BPM}
\label{sec:lcbpm}
In the main linac of the ILC, two types of cryomodules for beam acceleration will be used: the Type-A and the Type-B  
where the Type-A cryomodules consist of 1.3\,GHz nine 9-cell superconducting radio-frequency (RF) cavities in a row 
whereas the Type-B cryomodules substitute one of the 9-cell cavities in the center 
with a ``quadrupole package'' that contains a quadrupole magnet for beam 
focusing, dipole correctors for beam steering, and a cavity BPM for beam position monitoring \cite{ilc_tdr}.

To avoid the effects from the accelerating cavity mode at 1.3\,GHz and its higher order modes, the dipole mode (TM$_{110}$)
of the cavity BPM is designed to resonate at a higher frequency of 2.040\,GHz. To fit in the limited space inside
the quadrupole package, a re-entrant design is adopted for the L-band cavity BPM.

We have fabricated three L-band cavity BPMs as a prototype \cite{tupc097}. Due to fabrication tolerances, the resonant frequencies
of the TM$_{110}$ modes of the three BPMs are slightly different as 2.015\,MHz, 2.050\,MHz, and 2.020\,MHz 
(referred to as ``BPM-A'', ``BPM-B'', and ``BPM-C'', respectively). A low quality factor is favored for cavity BPMs to have 
a fast decay time of the
dipole mode signals so that the bunch to bunch signals are well separated and acquired. The loaded quality factors
of the fabricated BPMs are measured to be about 200 that yields a decay time of about 15\,ns,
which is reasonably short for the main linac of the ILC where the bunch spacing is 554\,ns.

The beam sizes in the main linac are designed to be 24-30\,nm in vertical and 8.4-9.4\,$\mu$m in horizontal.
To monitor the beam position in the main linac, it requires a BPM resolution better than 5\,$\mu$m, however,
we desire to achieve a few hundreds of nanometers of position resolution to have a sufficient margin for the beam-based alignment
and feedback control of the beam orbit.

\section{Experimental Setup}
\label{sec:experimental_setup}

The L-band cavity BPMs are installed in the Accelerator Test Facility (ATF) \cite{atf2} at the High Energy 
Accelerator Research Organization (KEK). 
We installed those at the end of the S-band linac of the ATF between
the bending magnets for the beam transport to the damping ring and the beam dump as shown in Fig. \ref{fig:bpm_station}.
There are two sets of dipole
magnets for the beam steering (each set consists of two magnets for horizontal and vertical steering.)
After the steering magnets, three L-band cavity BPMs are installed in a row. Before and after the BPMs, there 
are stripline BPMs for calibration purposes.

The pickup signal processing is illustrated in Fig. \ref{fig:rf_chain}.
Two opposite ports of each axis of the cavity BPM are used to pick up the dipole mode signals.
Since a dipole mode of a cavity BPM has an opposite phase between the two opposite ports, 
the signal picked up from the two ports are combined with a $180^{\circ}$ hybrid coupler to enhance the position sensitivity,
specifically $\Delta$-port of the hybrid coupler is used for the combination 
whereas $\Sigma$-port is terminated by 50\,$\Omega$ load.
Combined signals are downconverted to an intermediate frequency (IF) of 80\,MHz by a electronics we developed
(referred to as ``downconverter'').
Finally, the IF signals are digitized by two 4-channel oscilloscopes with 10-bit vertical resolution at a sampling 
rate of 2.5\,GS/s triggered by a trigger signal from the ATF timing system that is synchronized to the beam arrival time \cite{atf_llrf_timing} .

\begin{figure}[t]
\centerline{\includegraphics[width=3.5in]{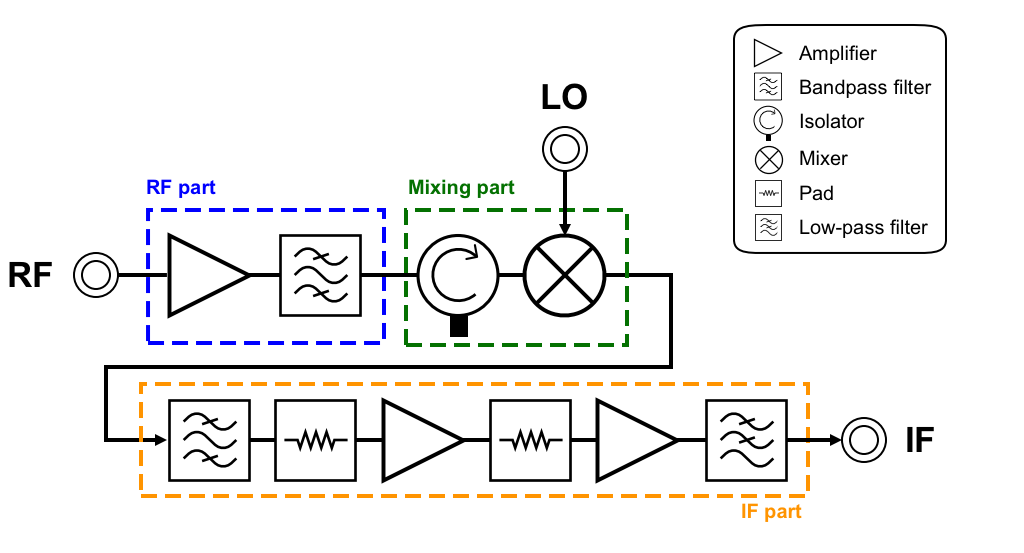}}
\caption{\label{fig:downconverter}
The downconverter used for the signal processing of the L-band cavity BPMs. One channel out of 8 identical channels is shown.}
\end{figure}

The downconverter used in the experiment provides 8 channels for the signal processing as shown in Fig. \ref{fig:downconverter}. 
It consists of three parts of
an RF part, a mixing part, and an IF part. All 8 channels are designed 
to be identical. The RF signal is amplified by an RF amplifier with a gain of about 21.3\,dB at 2040\,MHz, and filtered
by a band-pass filter with a bandwidth of 400\,MHz centered at 2040\,MHz and an insertion loss of about 3\,dB.
Before the downconversion by a mixer, any reflection from the mixer is removed by an isolator with an insertion loss
of about 1.5\,dB. The RF signal is downconverted to an IF of 80\,MHz by a mixer with a local oscillator (LO) signal 
provided by an external LO source described later with an conversion loss of about 7.3\,dB. The IF signal is further
filtered by a band-pass filter with a bandwidth of 20\,MHz centered at 80\,MHz and an insertion loss of about 2.5\,dB.
Then, it is amplified by two amplifiers with a gain of about 17\,dB each. The IF signal is finally filtered by a low-pass
filter with a cutoff frequency of 90\,MHz and an insertion loss of about 1\,dB. The net gain of the downconverter is 
about 25\,dB with a total noise figure of about 6.6\,dB.

The LO signal for the downconversion is provided by an external, custom-made LO source. The LO source is equipped with 
an integrated crystal oscillator fixed at 25\,MHz as a source clock. The source clock is multiplied to desired frequencies
by a voltage-controlled oscillator (VCO) and a phase-locked loop (PLL) circuit controlled by a computer through USB ports. 
The LO source is capable to generate LO signals
from 10\,MHz to 15\,GHz with output powers from 5 to 10\,dBm. As described in Sect. \ref{sec:lcbpm}, 
our BPMs have a slightly different resonant frequency of TM$_{110}$ modes, therefore, the LO source provides three different
LO signals at 1935\,MHz, 1970\,MHz, and 1940\,MHz for BPM-A, BPM-B, and BPM-C, respectively, by three different channels.
As mentioned, the power of the LO signal from the LO source is between 5 and 10\,dBm, and it is not enough to turn on 
the mixer inside the downconverter that requires at least 17\,dBm for the operation. Therefore, we place amplifiers
with a gain of about 28\,dB after the LO source. Since we need 6 LO signals for 
both axes of three BPMs, the LO signals from three channels of the LO source need to be splitted. Thus, we add three two-way
power splitters with a total loss of about 3.8\,dB, amplitude and phase unbalance of 0.05\,dB and 0.19$^{\circ}$, respectively.
Since the LO signal is generated from an internal clock of the LO source that is not synchronized to the beam,
the phase of the LO signal is not locked to the beam arrival time. This is a potential source of inaccurate measurements
as described in later section.

The control of steering magnets and the data acquisition from the magnets and the stripline BPMs are done with the EPICS \cite{epics}
database and the associated software tools already deployed at the ATF. The data acquisition from the oscilloscopes for 
the experiment is performed by a custom-made program developed in Python. The program stores the waveform data from the 
oscilloscopes and auxiliary data from the EPICS Process Variables (PVs) such as the beam charge, the current of the steering magnets, and the
readings of the stripline BPMs in an ASCII format with a laptop computer connected through ethernet for the offline analysis.

\section{Methodology}
\label{sec:methodology}
To measure the position resolution of the L-band cavity BPMs, we first obtain the correlation between the measured positions
of which we desire to find the resolution and of which are measured in other channels. In a matrix form, it can be written as
\begin{eqnarray}
    \mathbf{d}_{k} & = & \mathbf{D}_{\not{k}}\cdot\mathbf{v}
\end{eqnarray}
where $\mathbf{d}_{k}$ is a matrix of the measured positions in the channel $k$, $\mathbf{D}_{\not{k}}$ is a matrix of the 
measured positions in all other channels, and $\mathbf{v}$ is a vector of the correlation coefficients. The subscript $k$ indicates
the channel of which we want to find the resolution, and $\not{k}$ indicates all other channels except $k$. 
To obtain $\mathbf{v}$, we apply the singular value decomposition (SVD) method to $\mathbf{D}_{\not{k}}$ as
\begin{eqnarray}
    \mathbf{D}_{\not{k}} & = & \mathbf{U}\cdot\mathbf{S}\cdot\mathbf{V}^{T}
\end{eqnarray}
where $\mathbf{U}$ and $\mathbf{V}$ are orthogonal matrices and $\mathbf{S}$ is a diagonal matrix with singular values.
Then, the correlation coefficients can be obtained by
\begin{eqnarray}
    \mathbf{v} & = & \mathbf{D}^{-1}_{\not{k}}\cdot\mathbf{d}_{k} \nonumber \\
               & = & \mathbf{V}\cdot\mathbf{S}^{-1}\cdot\mathbf{U}^{T}\cdot\mathbf{d}_{k}.
\end{eqnarray}
Once we obtain the
correlation coefficients $\mathbf{v}$, one can find predicted positions by 
$\mathbf{d}^{\mathrm{pred.}}_{k}=\mathbf{D}_{\not{k}}\cdot\mathbf{v}$. Finally, we define a residual of the channel $k$,
\begin{eqnarray}
    \mathbf{R}_{k} & \equiv & \mathbf{d}^{\mathrm{pred.}}_{k}-\mathbf{d}_{k},
\end{eqnarray}
and the standard deviation of the residual distribution $\sigma_{\mathbf{R}_{k}}$ is the basis of the position resolution 
of the channel $k$.

In the experiment, what we measure with the BPMs are waveforms of signals in a unit of voltage in the time domain. The beam position
are extracted by integrating signals around the expected IF frequency in the frequency domain after a fast Fourier transform (FFT) 
of the waveforms. Therefore, in the above equations, $\mathbf{d}_{k}$ and $\mathbf{D}_{\not{k}}$ are matrices of the integrated 
signal values in a unit of arbitrary FFT amplitude that depends on a normalization scheme of the FFT. 
To convert the unit to a physical unit such as micrometers, one needs to calibrate the FFT amplitude referring to the beam position
measured by auxiliary BPMs such as stripline BPMs in our case. This calibration procedure is where the steering magnets and 
stripline BPMs come into play in the experiment. The calibration procedure finds a ratio between the FFT amplitude measured by
a cavity BPM and the beam position measured by a stripline BPM when the beam is steered across the cavity BPM by the steering magnets:
\begin{eqnarray}
    C_{k} & = & \frac{\Delta x_{k}}{\Delta A_{k}}
\end{eqnarray}
where $C_{k}$ is a calibration factor for the channel $k$, $\Delta x_{k}$ is a change of the beam position measured by
the stripline BPM, and $\Delta A_{k}$ is a change of the FFT amplitude measured by the cavity BPM. The calibration factor 
is applied to $\sigma_{\mathbf{R}_{k}}$ to convert the unit to micrometers.

\begin{figure*}[t]
\centerline{\includegraphics[width=7in]{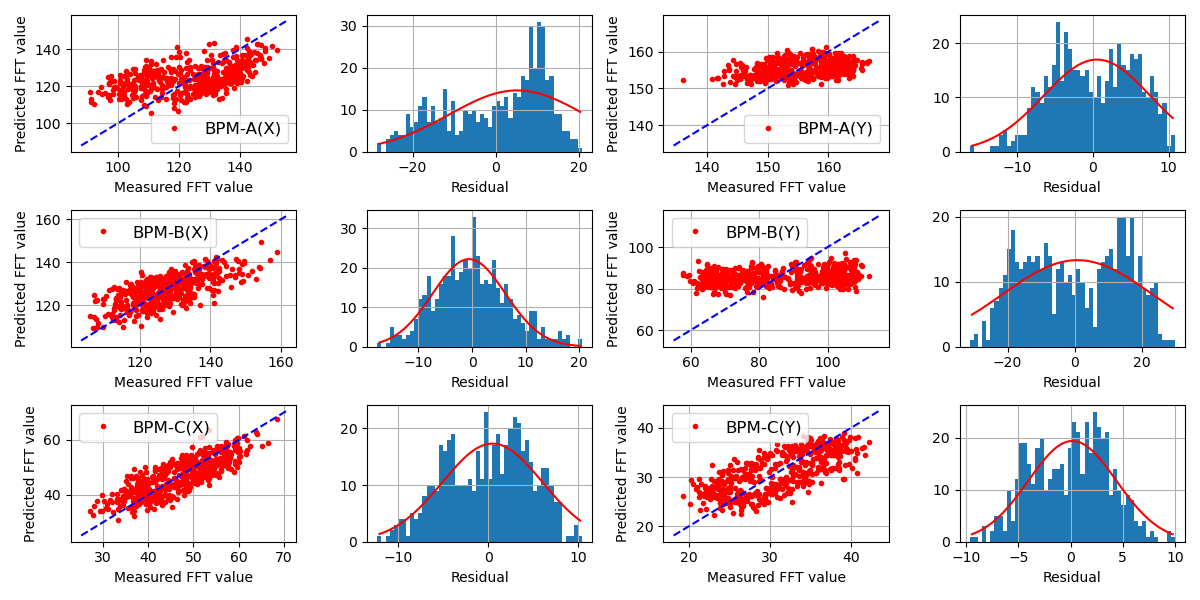}}
\caption{\label{fig:correlation_2024}
The correlation between the measured and predicted signals (scatter plot), and the residual distribution (histogram) from the 2024 experiment
for BPM-A horizontal (top left), BPM-A vertical (top right), BPM-B horizontal (middle left), BPM-B vertical (middle right),
BPM-C horizontal (bottom left), and BPM-C vertical (bottom right). FFT values and residuals are shown in arbitrary units throughout
the paper.
Blue dashed line in the scatter plots indicates the ideal correlation. Red line in the histograms is a fit with a Gaussian function.}
\end{figure*}

The converted values need to be corrected with considering the experimental setup such as an upstream attenuations and the
positional relation of the BPMs so-called geometrical factor. Due to a dynamic range of the signal processing chain, one may
need to add attenuators after the BPMs to avoid signal saturation when the beam charge is high. This upstream attenuation
reduces the signal amplitude, therefore, the position resolution is worsened by the same factor \cite{prst_11_062801}. 
The geometrical factor comes from the fact that the beam position at a BPM is not independent but correlated to the beam positions
at other BPMs due to the beam transport along the beamline. Therefore, the position resolution measured at a BPM needs to be corrected
by the geometrical factor,
\begin{eqnarray}
    G_{i} & = & \left[
        \left(\frac{s_{jk}}{s_{jk}}\right)^{2}
        +\left(\frac{s_{ik}}{s_{jk}}\right)^{2}
        +\left(\frac{s_{ij}}{s_{jk}}\right)^{2}
    \right]^{-\frac{1}{2}},
\end{eqnarray}
where $i$, $j$, and $k$ are the indices of three BPMs along the beamline, and $s_{ij}$ is the distance between BPM $i$ and BPM $j$
with $s_{ij}=s_{ji}$.

The position resolution of the channel $k$ with all corrections can be summarized as
\begin{eqnarray}
    \sigma_{k} & = & \sigma_{\mathbf{R}_{k}}\frac{G_{k}}{C_{k}}.
\end{eqnarray}

If a beam charge in an experiment is different from the nominal operating condition, one may claim a projected position resolution
by scaling the measured position resolution with the beam charge. The nominal beam charge of the ATF is 1.6\,nC \
($10^{10}$ electrons per bunch), however, our measurement was done with a lower beam charge as described in Sect.
\ref{sec:calibration_and_resolution}. This will be considered in our final results as well.

\section{Data Collection}
\label{sec:data_collection}
We conducted the experiments to measure position resolutions of the L-band cavity BPMs in 2024 and 2025. In the experiments,
we first took data for a calibration (referred to as a ``calibration run'') of the cavity BPMs by steering the beam position with 
the steering magnets across the cavity BPMs. After a calibration run, we took data for position resolution measurements 
(referred to as a ``resolution run'') with the beam kept at a fixed position without any steering. Both calibration and resolution
runs consist of multiple measurements that each contains single waveforms from all six channels of the cavity BPMs.
Each waveform was taken at a sampling rate of 2.5\,GS/s for a duration of about 1.2\,$\mu$s that contains
about 3000 samples.

For the signal amplitudes in the frequency domain, we first found the peak frequency around the expected IF of 80\,MHz
for each channel from averaged FFT amplitudes over multiple measurements. Then, we integrated the FFT amplitudes over
a bandwidth of about 8\,MHz centered at the peak frequency to obtain the signal amplitude for each measurement.

The 2024 experiment was performed in November 2024, and we took 50 measurements for the calibration purpose 
over 9 beam positions in horizontal and 11 beam
positions in vertical for each, thus 1000 waveforms in total were taken. The resolution run followed
the calibration run, and we took 500 waveforms with the beam kept at a fixed position.

In the 2025 experiment that was performed in May 2025, two different configurations 
were applied due to the reasons described in Sect. \ref{sec:gain_and_phase}. 
 The calibration runs took 50 measurements over 7 beam positions in both horizontal and vertical for each, 
and the resolution runs took 1000 measurements at a fixed beam position.

\begin{figure*}[t]
\centerline{\includegraphics[width=7in]{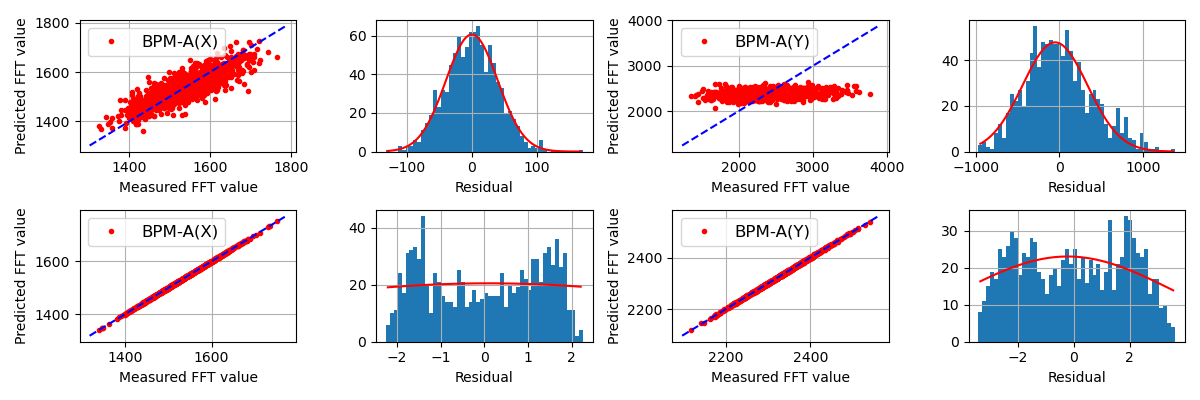}}
\caption{\label{fig:correlation_simulation}
Simulated correlations between the measured and predicted signals (scatter plot), and the residual distribution (histogram). 
The upper figures are from the simulation with the gain incoherency of which was measured as Fig. \ref{fig:gain_fluctuation}
whereas the lower figures are from the simulation with the phase incoherency of $2\pi$. Only BPM-A cases are shown here for brevity.
Blue dashed line in the scatter plots and red line in the histograms have the same meanings as in Fig. \ref{fig:correlation_2024}.}
\end{figure*}

\section{Effect of Gain and Phase Incoherency}
\label{sec:gain_and_phase}
In the 2024 experiment, we found the correlations between predicted and measured positions were badly broken as shown in Fig. 
\ref{fig:correlation_2024}. Ideally, the correlation should be aligned along the diagonal line in a scatter plot of the predicted
and the measured positions. The 2024 data shows two problems: (1) the correlation is flat along the measured position axis,
meaning that the predicted position was not estimated well, and (2) there are two clusters in the correlation plot that yield
two peaks in the residual distribution. These two problems indicate that there were problems in the signal processing chain,
and the measurements were not quite reliable.

An obvious suspect is the unlocked phase of the LO signal to the beam arrival time as described in Sect. \ref{sec:experimental_setup}.
Another suspect is the gain incoherency among the channels of the downconverter. To confirm these effects, we performed simulations 
to find the qualitative effect of the gain and the phase incoherency on the position resolution.

The simulation generates synthetic waveforms in the time domain for six channels of the cavity BPMs with a given beam position as:
\begin{eqnarray}
    V(t) & = & GV_{0}e^{-t/2\tau}\left(1-e^{-t/2\tau_{\mathrm{rise}}}\right)\cos(\omega t+\phi)
\end{eqnarray}
where $G$ is the signal gain, $V_{0}$ is the amplitude proportional to the beam offset, $\tau$ is the decay time of the cavity, 
$\tau_{\mathrm{rise}}$ is the rise time of the signal due to the RF chain response, $\omega$ is the angular frequency of 
the TM$_{110}$ mode of the cavity, and $\phi$ is the phase offset. The signal amplitude $V_{0}$ is given as
\begin{eqnarray}
    V_{0} & = & \frac{q\omega^{2}}{2}\sqrt{\frac{Z}{Q_{\mathrm{ext}}}\frac{R}{Q}}e^{-\frac{\omega^{2}\sigma_{z}^{2}}{2c^{2}}}
\end{eqnarray}
where $q$ is the bunch charge, $Z$ is the load impedance (=50 $\Omega$), $Q_{\mathrm{ext}}$ is the external quality factor of 
the external coupling, $R/Q$ is the shunt impedance over quality factor of TM$_{110}$ mode of the cavity, $\sigma_{z}$ is 
the bunch length, and $c$ is the speed of light. The beam position is encoded in $R/Q$ of the TM$_{110}$ mode of a cylindrical
cavity as
\begin{eqnarray}
    \frac{R}{Q} & = & 1.873\times10^{-24}\cdot LT^{2}x^{2}\omega^{3}
\end{eqnarray}
where $L$ and $T$ are the length and the transit time factor of the cavity, and $x$ is the beam offset from the center of the cavity.

For the tests, we examined two cases separately: fluctuating $G$ of channels independently with a Gaussian distribution 
of a given standard deviation, and fluctuating $\phi$ of channels independently with a uniform distribution between 0 and $2\pi$.
Figure \ref{fig:correlation_simulation} shows the correlation plots from the simulations of the cases. As shown, we confirm that
the gain incoherency among the channels yields a similar broken correlation plot as the 2024 data 
whereas the phase incoherency yields two peaks in the residual distribution. 

\begin{figure}[t]
\centerline{\includegraphics[width=3.5in]{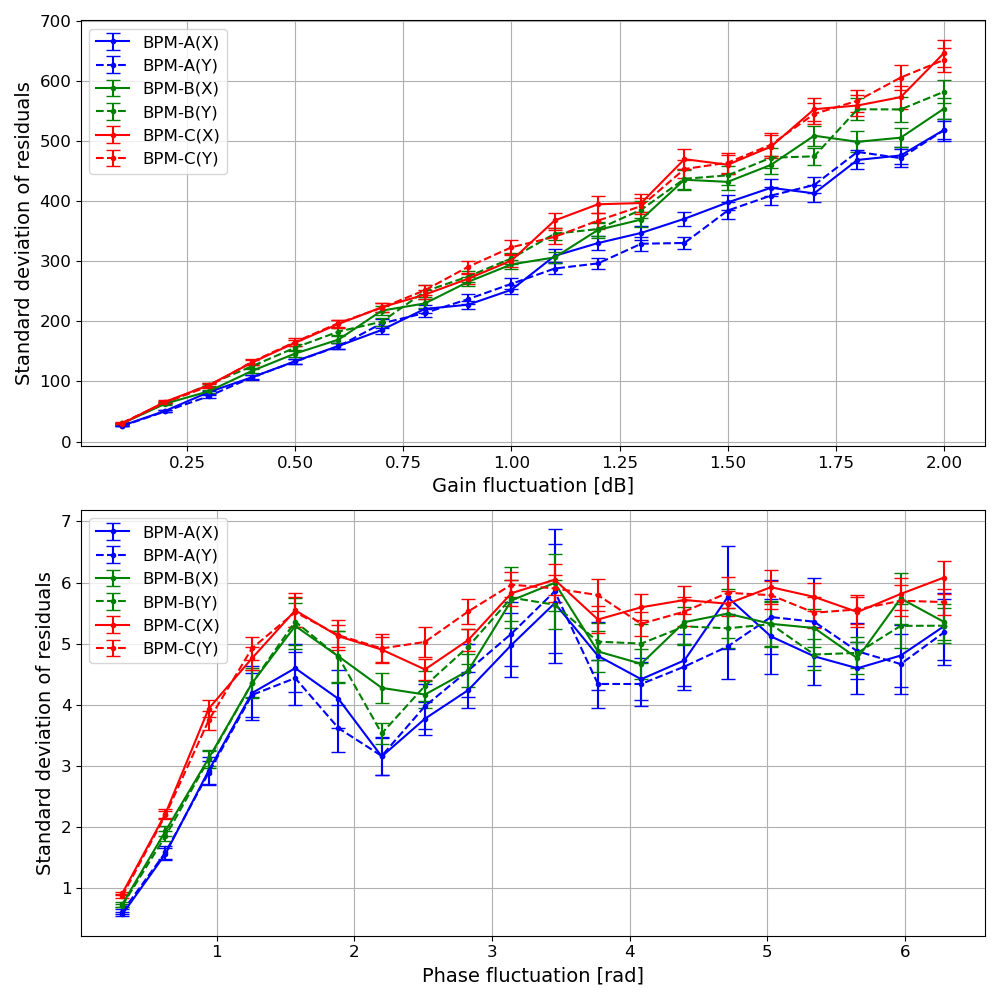}}
\caption{\label{fig:residual_simulation}
Standard deviation of the residual distribution ($\sigma_{\mathbf{R}_{k}}$) from the simulations with different levels of gain and phase incoherency.}
\end{figure}

Figure \ref{fig:residual_simulation} shows how $\sigma_{\mathbf{R}_{k}}$, equivalently the position
resolution, changes with the gain and phase incoherency. Since the phase incoherency
yields two peaks in the residual distribution, it degrades the position resolution. On the other hand, it reasonably maintains 
the linearity of the correlation between the predicted and the measured positions, therefore, the position resolution 
is saturated after a certain level of the phase incoherency, around 1.5\,rad in this simulation. The gain incoherency, however,
does not maintain the linearity of the correlation, and makes the residual distribution broader. Therefore, the position resolution
continuously degrades as the gain incoherency increases.

\begin{figure}[t]
\centerline{\includegraphics[width=3.5in]{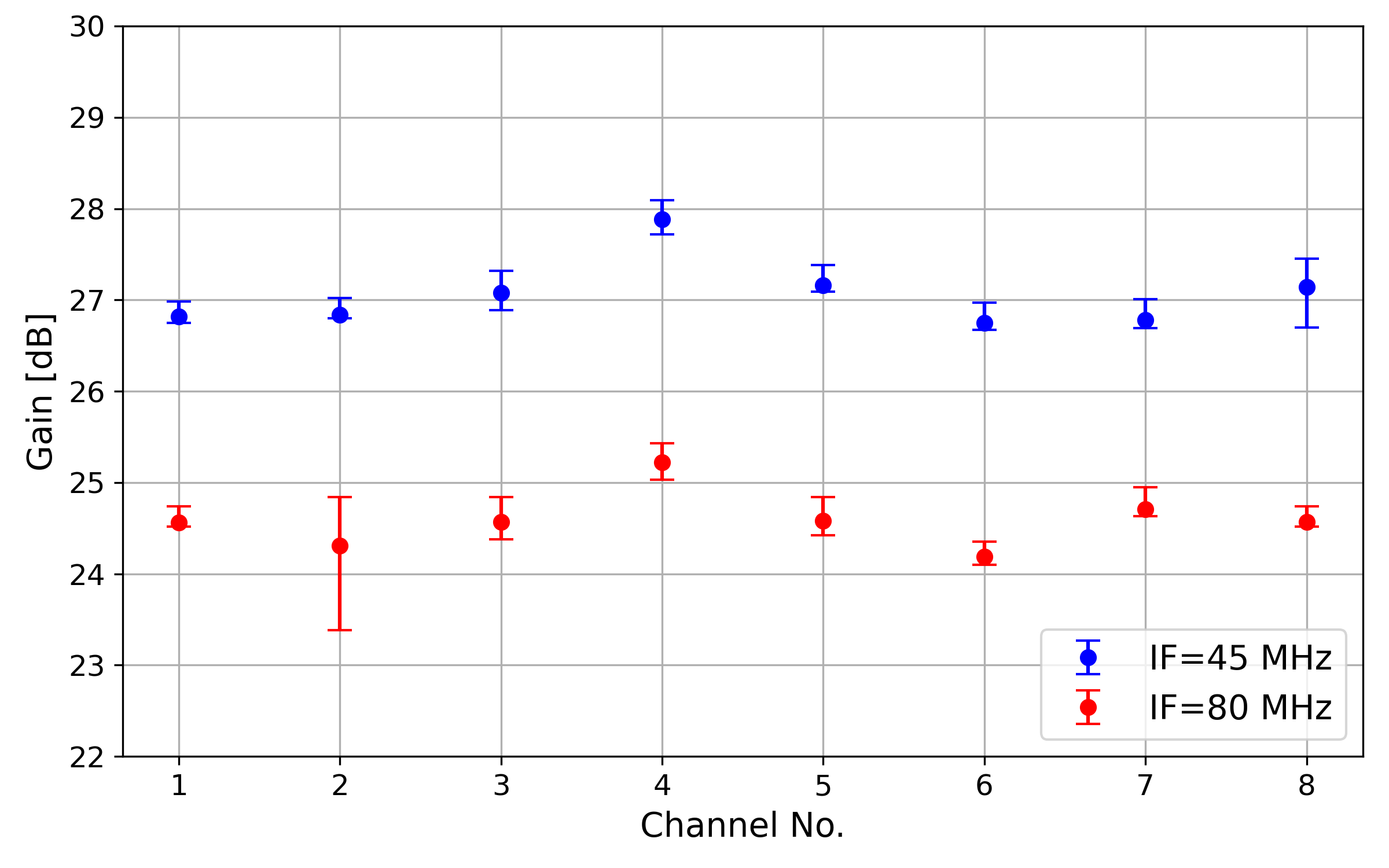}}
\caption{\label{fig:gain_fluctuation}
Measured gains of the downconverter for different IF of 45\,MHz (blue dots) and 80\,MHz (red dots). The vertical error bars were obtained by taking
maximum and minimum values over time.}
\end{figure}

The gain incoherency was confirmed by measuring the gain of the downconverter as shown in Fig. \ref{fig:gain_fluctuation}.
To measure the gain fluctuation, we provided a continuous wave (CW) RF signal from a signal generator to the RF input of
the downconverter, and measured the IF output signal power with a spectrum analyzer over time. To downconvert the CW RF signal 
to 80\,MHz IF signal, we provided a LO signal from our LO source to the LO input of the downconverter. 
As shown in Fig. \ref{fig:gain_fluctuation}, all channels except for channel 2 show reasonable gain stability of less than 0.5\,dB
peak-to-peak over time. However, channel 2 shows a significant gain fluctuation of about 1.5\,dB peak-to-peak over time.

During the test, we found that the band-pass filters after mixers were not working properly for all channels. This allowed us 
to try another IF frequencies by changing the LO frequencies. For example, as shown in Fig. \ref{fig:gain_fluctuation}, 
IF of 45\,MHz shows a better RF characteristics.

Based on these findings, we constructed two different configurations for the 2025 experiment: one with the same IF of 45\,MHz
to minimize the effect of the gain incoherency, and the other with a common LO signal to get rid of the effect of the phase 
incoherency. For the latter configuration, we provided a common LO signal at 1970\,MHz to all three BPMs, therefore, 
the IF frequencies fall into different IF of 45\,MHz, 80\,MHz, and 50\,MHz for BPM-A, BPM-B, and BPM-C, respectively.

\section{Calibration and Position Resolution Measurement}
\label{sec:calibration_and_resolution}
As described in Sect. \ref{sec:data_collection}, we collected calibration data with steering the beam across the cavity BPMs.
To ensure the same beam conditions between the calibration and the resolution runs, the calibration run was performed right before
the resolution run, without changing any machine parameters.

\begin{figure}[t]
\centerline{\includegraphics[width=3.5in]{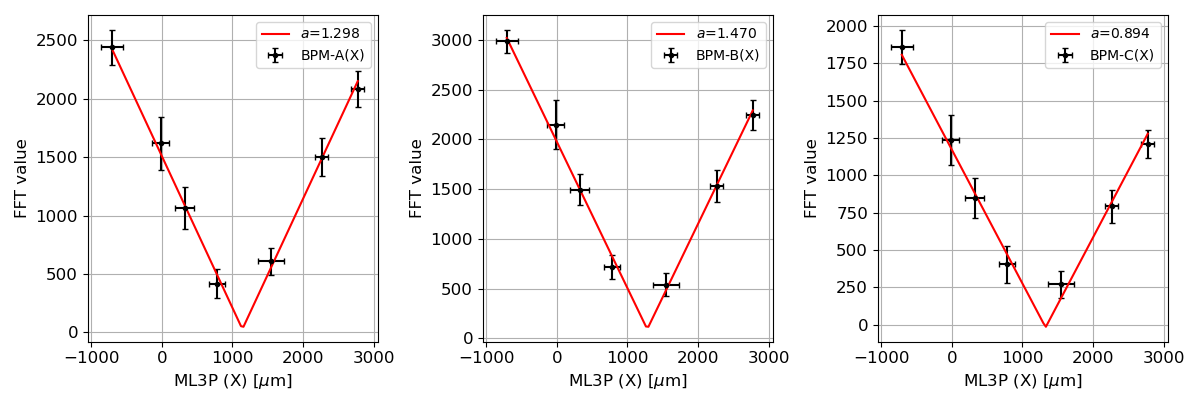}}
\centerline{\includegraphics[width=3.5in]{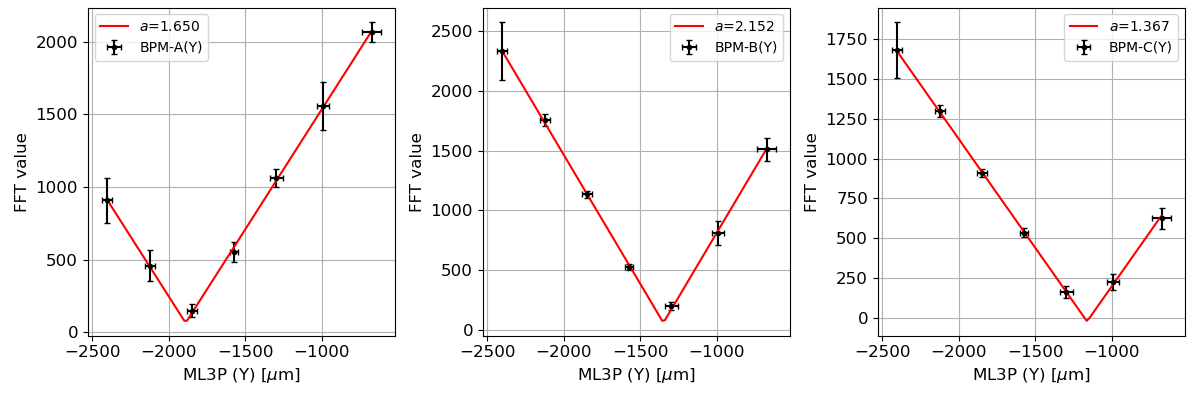}}
\caption{The ``V-shaped'' calibration curves for horizontal (top) and vertical (bottom) 
of the BPM-A (left), BPM-B (middle), and BPM-C (right) 
obtained from the same IF configuration in the 2025 experiment. Fits by the linear function of $y=a|x-b|+c$ are shown as red lines.}
\label{fig:calibration}
\end{figure}

To obtain the calibration data, we first found the beam position 
near the center of the cavity BPMs by adjusting the steering magnets. From this position, we chose seven beam positions
around the center in both horizontal and vertical for the same IF and the common LO configurations. As shown 
in Fig. \ref{fig:calibration}, we obtained a ``V-shaped''
calibration curve by plotting the FFT amplitude measured by the cavity BPM against the beam position measured by the stripline BPM.
Then, the ``V-shaped'' curve is fitted by a linear function of $y=a|x-b|+c$ to reflect the fact that the signal amplitude
measured by the cavity BPM is blind to the sign of the beam position. As described in Sect. \ref{sec:methodology}, the fit parameter
$a$ (slope) is the calibration factor, $C_{k}$. Table \ref{table:calibration_factors}
summarizes the calibration factors obtained from the two configurations.

\begin{table}
\caption{Calibration factors obtained from the 2025 experiment.}
\label{table:calibration_factors}
\begin{tabular}{|p{55pt}|p{80pt}|p{80pt}|}
\multicolumn{3}{p{215pt}}{(a) Calibration factors from the same IF configuration}\\
\hline
& Horizontal [$\mu\mathrm{m}^{-1}$] & Vertical [$\mu\mathrm{m}^{-1}$] \\
\hline
BPM-A & $1.298\pm0.054$ & $1.650\pm0.023$ \\
BPM-B & $1.470\pm0.072$ & $2.152\pm0.032$ \\
BPM-C & $0.894\pm0.056$ & $1.367\pm0.016$ \\
\hline
\multicolumn{3}{p{215pt}}{* The uncertainties are fit errors.}\\
\multicolumn{3}{p{215pt}}{}\\
\multicolumn{3}{p{215pt}}{(b) Calibration factors from the common LO configuration}\\
\hline
& Horizontal [$\mu\mathrm{m}^{-1}$] & Vertical [$\mu\mathrm{m}^{-1}$] \\
\hline
BPM-A & $0.442\pm0.050$ & $0.598\pm0.051$ \\
BPM-B & $0.449\pm0.026$ & $0.486\pm0.034$ \\
BPM-C & $0.492\pm0.036$ & $0.652\pm0.050$ \\
\hline
\multicolumn{3}{p{200pt}}{* The uncertainties are fit errors.}
\end{tabular}
\end{table}

For the resolution measurement, we chose a beam position where the beam passed through the stripline BPM position 
of about 2500\,$\mu$m in horizontal and -2500\,$\mu$m in vertical for the same IF configuration, and 
about 3600\,$\mu$m in horizontal and -3100\,$\mu$m in vertical for the common LO configuration. The beam charge
was measured about 0.96\,nC and 0.86\,nC for the same IF and the common LO configurations, respectively.

\begin{figure*}[t]
\centerline{\includegraphics[width=7in]{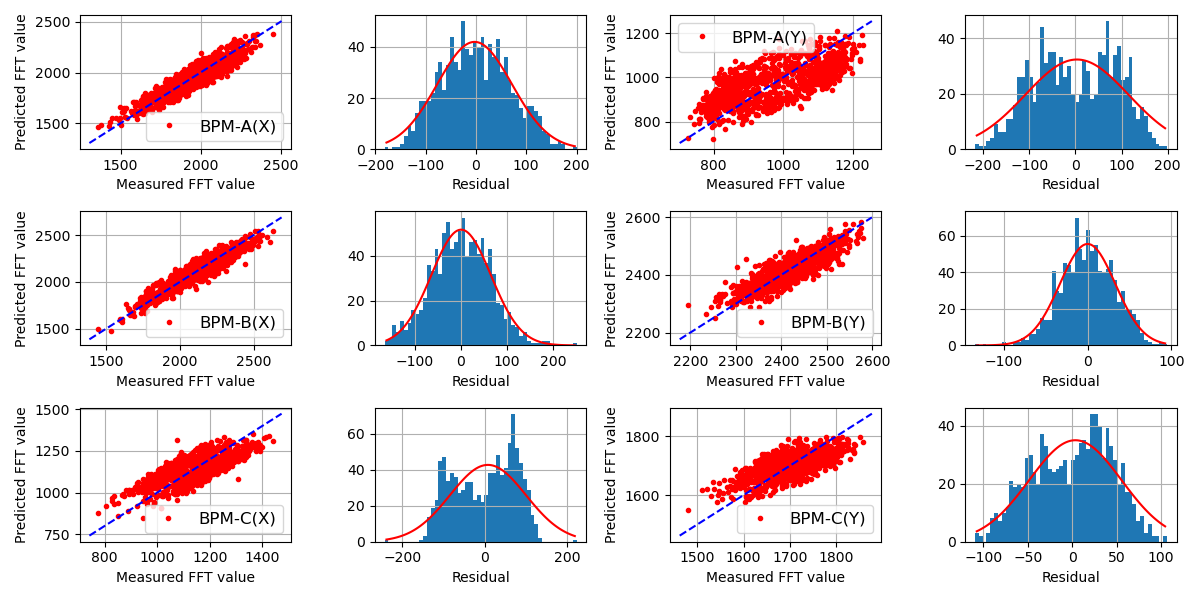}}
\caption{\label{fig:resolution_2025a1}
The correlation between the measured and predicted signals (scatter plot), and the residual distribution (histogram) 
from the same IF configurations in the 2025 experiment
for BPM-A horizontal (top left), BPM-A vertical (top right), BPM-B horizontal (middle left), BPM-B vertical (middle right),
BPM-C horizontal (bottom left), and BPM-C vertical (bottom right). 
Blue dashed line in the scatter plots and red line in the histograms have the same meanings as in Fig. \ref{fig:correlation_2024}.}
\end{figure*}

Figure \ref{fig:resolution_2025a1} shows the scatter plot of the predicted and the measured positions along with the residual
distribution for the same IF configuration in the 2025 experiment. As shown, the measurements are well populated along
the diagonal line in the correlation plot and this validates the gain incoherency in the 2024 experiment. 
There are still two peaks in the residual distribution since the same IF configuration
suffers from the phase incoherency as discussed in Sect. \ref{sec:gain_and_phase}. Therefore, the resolutions of some channels
are slightly worse than others as summarized in Table \ref{table:position_resolutions} (a).

\begin{figure*}[bt]
\centerline{\includegraphics[width=7in]{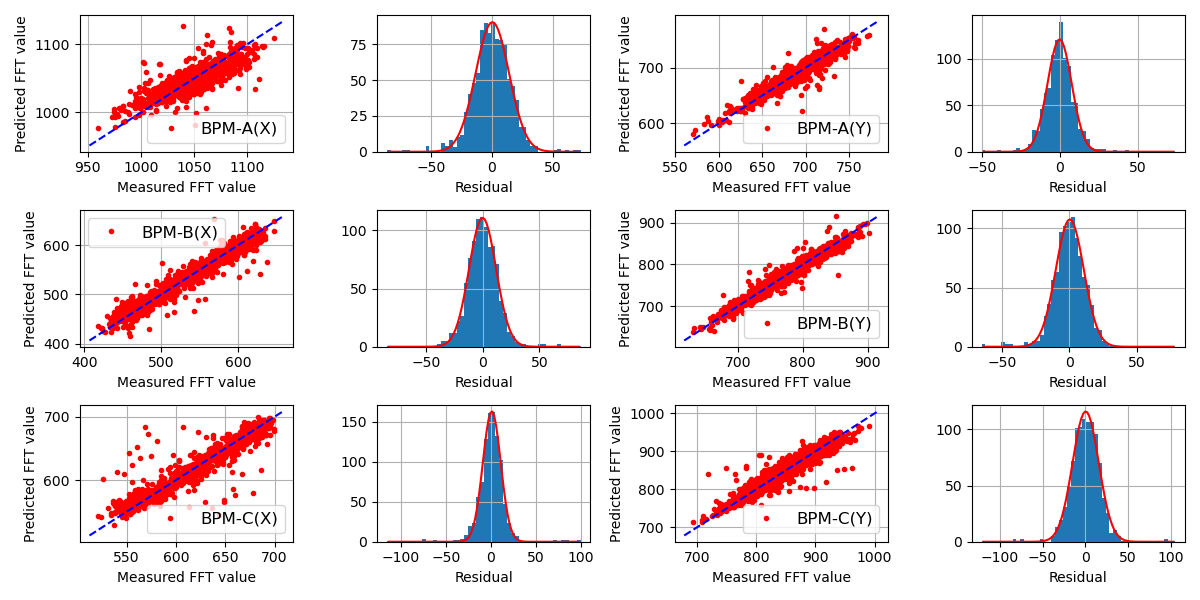}}
\caption{\label{fig:resolution_2025a2}
The correlation between the measured and predicted signals (scatter plot), and the residual distribution (histogram) 
from the common LO configurations in the 2025 experiment
for BPM-A horizontal (top left), BPM-A vertical (top right), BPM-B horizontal (middle left), BPM-B vertical (middle right),
BPM-C horizontal (bottom left), and BPM-C vertical (bottom right). 
Blue dashed line in the scatter plots and red line in the histograms have the same meanings as in Fig. \ref{fig:correlation_2024}.}
\end{figure*}

The result of the common LO configuration is shown in Fig. \ref{fig:resolution_2025a2}. As shown, the correlation plot
is well aligned along the diagonal line, and the residual distribution shows no two-peak structure. This validates
the effect of the phase incoherency in the 2024 experiment. As summarized in Table \ref{table:position_resolutions} (b),
the result with the common LO configuration is better and more reliable compared to the same IF configuration.

For both configurations, we added 20\,dB attenuators after the cavity BPMs to avoid signal saturation in the downconverter.
Therefore, the position resolutions need to be corrected by a factor of 10 due to the upstream attenuation as described
in Sect. \ref{sec:methodology}. The geometrical factors for BPM-A, BPM-B, and BPM-C are calculated to be 0.450, 0.815, and 0.365,
respectively. After all corrections, the projected position resolutions assuming a beam charge of 1.6\,nC are summarized 
in Table \ref{table:position_resolutions}.

\begin{table}
\caption{Projected position resolutions of the L-band cavity BPMs obtained from the 2025 experiments.}
\label{table:position_resolutions}
\begin{tabular}{|p{55pt}|p{80pt}|p{80pt}|}
\multicolumn{3}{p{215pt}}{(a) The same IF configuration}\\
\hline
& Horizontal [nm] & Vertical [nm] \\
\hline
BPM-A & $1552\pm114$ & $1843\pm182$ \\
BPM-B & $2165\pm164$ & $756\pm47$ \\
BPM-C & $2277\pm284$ & $838\pm65$ \\
\hline
\multicolumn{3}{p{200pt}}{* The uncertainties are statistical.}\\
\multicolumn{3}{p{215pt}}{}\\
\multicolumn{3}{p{215pt}}{(b) The common LO configuration}\\
\hline
& Horizontal [nm] & Vertical [nm] \\
\hline
BPM-A & $768\pm96$ & $319\pm33$ \\
BPM-B & $1188\pm93$ & $951\pm83$ \\
BPM-C & $404\pm36$ & $469\pm44$ \\
\hline
\multicolumn{3}{p{200pt}}{* The uncertainties are statistical.}
\end{tabular}
\end{table}

\section{Discussions}
As shown in Fig. \ref{fig:resolution_2025a1}, the same IF configuration improves the linearity of the correlation
compared to the 2024 experiment, however, the two-peak structure in the residual distribution still remains. This
confirms our discussion in Sect. \ref{sec:gain_and_phase} that the gain incoherency breaks the linearity of the correlation
whereas the phase incoherency yields the two-peak structure in the residual distribution. Figure \ref{fig:resolution_2025a2}
supports this discussion more concretely since the common LO configuration gets rid of the two-peak structure in the residual 
distribution. Since the common LO used in the 2025 experiment was not actually locked to the beam arrival time, there is still
some phase incoherency due to the beam timing jitter, and this needs to be improved in the future experiments.

As shown in Table \ref{table:position_resolutions} (b), the projected position resolutions of the BPM-B are slightly worse than
the others. We suspect that this is due to the poor behavior of the downconverter at the IF of 80\,MHz though we carefully
chose channels which seem more stable than others. Making stable channels of the downconverter is another thing 
to be improved in the future experiments.

Furthermore, the signal amplitudes acquired in our experiments are limited by the vertical resolution of the oscilloscopes
(10 bits). To fully exploit the performance of the L-band cavity BPMs, a higher vertical resolution of at least 12 bits is desired.

\section{Conclusion}
\label{sec:conclusion}
We have developed a prototype of an L-band cavity BPM for future linear colliders especially for the main linac of the ILC. 
To evaluate the performance 
of the BPMs, we have conducted beam tests at the ATF in KEK. In the 2024 experiment, we found that the gain and phase incoherency
among the channels of the downconverter significantly degraded the position resolution measurements. Based on the findings,
we conducted another experiment in 2025 with two different configurations to minimize the effects of the gain and phase incoherency.
As a result, we have successfully measured the projected position resolutions of the L-band cavity BPMs to be
about 300\,nm at best with a beam charge of 1.6\,nC.
For future works, we will improve the LO source to be locked to the beam arrival time to minimize the phase incoherency, 
will stabilize the downconverter to minimize the gain incoherency,
and will develop a dedicated data acquisition system with a higher vertical resolution. With these improvements, we
expect to achieve better position resolutions in the future experiments.

\end{document}